\documentclass[color,twocolumn,prl,aps,footinbib,superscriptaddress]{revtex4} 
\usepackage{graphics,graphicx,amssymb}

\newcommand{\newc}{\newcommand}
\newc\eg{{\it {e.g.}}}	\newc\vs{{\it {vs.}}}	\newc\etal{{\it {et al.}}}

\newc{\mhalf}{m_{1/2}}      \newc{\mzero}{m_0}

\newc\bsgamma{b\rightarrow s\gamma}
\newc\brbsgamma{{\mathcal B}(B\rightarrow X_s\gamma)}

\newc{\tanb}{\tan\beta}
\newc{\azero}{A_0}
\newc{\at}{A_t} \newc{\abot}{A_b} \newc{\atau}{A_\tau}
\newc{\bmu}{B\mu}           \newc{\sgn}{{\rm sgn}}
\newc{\mone}{M_1}           \newc{\mtwo}{M_2}

\newc{\bino}{\widetilde B}              \newc{\wino}{\widetilde W_3}
\newc{\higgsinob}{{\widetilde H}^0_b}   \newc{\higgsinot}{{\widetilde H}^0_t}

\newc{\mb}{m_b} \newc{\ms}{m_s}

\newc{\mw}{m_{W}}
\newc\msusy{M_{\rm SUSY}}
\newc{\mplanck}{M_{\rm P}}

\newc{\mub}{\mu_{b}}	\newc{\muw}{\mu_{W}}
\newc{\mususy}{\mu_{\rm SUSY}}

\newc{\Ci}{C_i}	\newc{\Cip}{C_i^{\prime}}

\newc{\deltadll}{\delta^d_{LL}}	\newc{\deltadlr}{\delta^d_{LR}}
\newc{\deltadrl}{\delta^d_{RL}}	\newc{\deltadrr}{\delta^d_{RR}}

\newc{\abund}{\Omega h^2}
\newc{\abundchi}{\Omega_\chi h^2}
\newc{\rhocrit}{\rho_{crit}}
\newc{\rhochi}{\rho_{\chi}}

\newc{\xf}{x_f}
\newc{\jxf}{J({\xf})}
\newc{\VEV}[1]{\langle #1 \rangle}

\newcommand\tev{\,\mbox{TeV}}
\newcommand\gev{\,\mbox{GeV}}

\newc{\ra}{\rightarrow}
\newc{\beq}{\begin{equation}}
\newc{\eeq}{\end{equation}}
\newc{\bea}{\begin{eqnarray}}
\newc{\eea}{\end{eqnarray}}

\newcommand\lsim{\mathrel{\rlap{\lower4pt\hbox{\hskip1pt$\sim$}}
    \raise1pt\hbox{$<$}}}
\newcommand\gsim{\mathrel{\rlap{\lower4pt\hbox{\hskip1pt$\sim$}}
    \raise1pt\hbox{$>$}}}

\newcommand{\Rn}[1]{{\uppercase\expandafter{\romannumeral#1}}}

\newcommand{\wt}[1]{\widetilde{#1}}

\newcommand{\chargino}{\chi^{\pm}} 
\newcommand{\charginom}{\chi^{-}} 

\newcommand{\neut}{\chi^0}
\newcommand\gluino{\widetilde{g}} \newcommand\mgluino{m_{\gluino}}

\newcommand{\sbottom}{\wt{b}} 
\newcommand{\sstrange}{\wt{s}}
\newcommand{\msquark}{m_{\wt{q}}}

\begin{document}
\draft
 \preprint{
}

\title{Weakened Constraints from \boldmath{$b\to s\gamma$} on
  Supersymmetry Flavor Mixing  Due\\ to Next--To--Leading--Order Corrections}

\author{Ken-ichi Okumura}
\altaffiliation
{Previously at Lancaster University.}
\affiliation{Department of Physics, KAIST, Daejeon, 305-701, Korea}
\author{Leszek Roszkowski}
\altaffiliation
{Previously at Lancaster University.}
\affiliation{Department of Physics and Astronomy, University of Sheffield, 
Sheffield, UK}



\begin{abstract}

We examine the process $B \to X_s \gamma$ in minimal supersymmetry
(SUSY) with general squark flavor mixings.  We include
all relevant next--to--leading order (NLO) QCD corrections and
dominant NLO SUSY effects from the gluino.
We find that gluino--squark corrections to down--type quark masses 
induce large NLO corrections to the dominant Wilson
coefficients whose size is often similar to those at LO,
 especially at large $\tanb$.
For $\mu>0$, destructive interference, and suppression by the
renormalization group running 
lead to a ``focusing effect'' of reducing 
the size of 
gluino corrections to the branching ratio, and also of
reducing the LO sensitivity to flavor mixings among squarks.  
 Constraints from $\brbsgamma$ on the SUSY--breaking
scale 
can become significantly weakened relative to the minimal flavor
violation case, even, at large $\tanb$, 
for small flavor mixings. 
The case of $\mu<0$ can also becomes allowed.
\end{abstract}


\maketitle

{\it Introduction}.---
The radiative decay $B \to X_s \gamma$ provides a powerful tool for
testing various extensions of the standard model (SM), such as
supersymmetry. This is because in $\bsgamma$ ``new physics''effects
can appear at a one--loop level, which is the same as the lowest--order SM
flavor changing neutral current (FCNC), and can therefore be of
comparable size~\cite{bbmr90}.  However, as both experimental
results~\cite{exp} and a SM prediction at the full 
next--to--leading--order (NLO) QCD
level~\cite{gm01,bcmu02} have reached an error of $\sim10\%$,
\begin{eqnarray}
\label{bsgexptvalue:ref}
{\cal B}(B \to X_s\gamma)_{\rm exp} &=&(3.34 \pm0.38)\times10^{-4},\\
{\cal B}(B \to X_s\gamma)_{\rm SM} &=&(3.70 \pm0.30)\times10^{-4},
\label{bsgsmvalue:ref}
\end{eqnarray}
it has become clear that 
little room is now left for
new physics contributions.  This has in turn been used to
impose severe constraints on flavor--violating (FV) interactions beyond
the SM. These are usually related to some scale $\Lambda$ at which new
physics appears. In the case of low--energy supersymmetry (SUSY),
 $\Lambda$ is set
by the SUSY--breaking scale $\msusy$, which is expected to remain
within ${\cal O}(1\tev)$ on the grounds of naturalness.

The SM contribution to $\bsgamma$ involves a one--loop exchange of the
top quark and the $W^{-}$. 
In 2--Higgs doublet models (2HDM), 
there is, in addition, a constructive contribution from the charged Higgs
$H^{-}$. 
A complete NLO analysis  in the SM has been performed in
several stages and recently completed in~\cite{bcmu02} and, likewise,
in the 2HDM~\cite{bg98,crs98,cdgg98}.

In SUSY, additional one--loop diagrams come from
an exchange of the charged Higgs--top $H^{-}$--$t$ and the
chargino--stop $\charginom$--$\wt{t}$.  The latter contributes
constructively (destructively), depending on whether the
Higgs/higgsino mass parameter $\mu>0$ or $<0$, 
 respectively.  Detailed studies of SUSY contributions to
$\brbsgamma$ have been performed beyond the
LO~\cite{cdgg98susy,bmu00,cgnw00,dgg00}. In particular,
dominant NLO contributions have been calculated~\cite{cgnw00,dgg00}
and shown to be important. They 
are
enhanced by factors involving large $\tanb=\langle
H^0_2\rangle/\langle H^0_1\rangle$ [the ratio of the neutral Higgs 
VEVs] and $\log\left(\msusy/\mw\right)$.

The good agreement between the
ranges~(\ref{bsgexptvalue:ref}) and (\ref{bsgsmvalue:ref}) has also been used
to derive stringent constraints on the mass spectra of the
superpartners in specific popular models, like the minimal
supersymmetric SM (MSSM), or the constrained MSSM (CMSSM).  For
example, in the CMSSM, the case of $\mu<0$ has been shown to be ruled
out, except for the very large common gaugino $\mhalf$ and scalar $\mzero$
masses ($\sim$ few~$\tev$), where SUSY contributions become tiny. 
For $\mu>0$, stringent lower bounds of a few
hundred~$\gev$ have also been derived on $\mhalf$ and $\mzero$,
especially at large $\tanb$~\cite{cmssmbsgamma}.

Such bounds, even if model dependent, have clear implications for
searches for SUSY in accelerators, for the neutralino as a dark matter
weakly interacting massive particle (WIMP),
and in other nonaccelerator processes.  For example, in dark matter
searches the constraint from $\bsgamma$ often forbids larger values
of the spin--independent scattering cross section of WIMPs on proton
$\sigma^{SI}_p$ that would otherwise be accessible to today's
experimental sensitivity~\cite{knrr1}.  

The above conclusions have often been reached by assuming, sometimes
implicitly, that the mixings among the mass eigenstates of
squarks 
closely resemble the Cabibbo--Kobayashi--Maskawa (CKM)
structure of the quark sector. This scenario, often called minimal
flavor violation (MFV), 
is perhaps the simplest way
of addressing the nagging flavor problem of SUSY.  FV interactions are
in general not forbidden by gauge symmetry and, without assuming any
organizing principle, can lead to exceeding experimental bounds by
many orders of magnitude~\cite{agis02}.  The mixings of the first
two generations of squarks are strongly constrained by the $K^0-\bar
K^0$ system but bounds on mixings with the third generation are
much weaker. In the case of $\bsgamma$, rather stringent constraints
have been derived, in the LO approximation, on the magnitude of
possible FV terms beyond MFV in specific scenarios for flavor
structure, but shown to be highly model dependent~\cite{bghw99,ekrww02}.
The question thus arises
about the robustness of constraints on SUSY derived in MFV or other specific scenarios.

In this Letter, we examine the process $\bsgamma$ in 
the MSSM with {\em general} flavor mixing (GFM) in the down--type
squark sector. In addition to 
the full NLO QCD corrections from the
SM+2HDM~\cite{bcmu02,bg98,crs98,cdgg98}, we introduce NLO QCD
resummation above $\mw$ and threshold corrections from the heavy
gluino field $\gluino$ 
in the presence of all mixings in the down--squark sector.
In contrast to MFV~\cite{cgnw00,dgg00}, most
analyses of the GFM have been done with only LO matching
conditions (MCs) for SUSY contributions~\cite{bghw99}, in which case 
NLO effects become an issue. In order to be able to compare
with the SM+SUSY contributions to $\brbsgamma$ in MFV and in LO, 
we calculate all $\tanb$--enhanced NLO contributions in the framework of GFM
and show that they lead to dramatic implications.
\vspace{0.15cm}
{\it Procedure}.---
The contributions to $\bsgamma$ from all states that are much heavier
than $\mb$ are described by the effective Lagrangian~\cite{gm01}
\beq
{\cal L} = \frac{4 G_F}{\sqrt{2}} V_{ts}^*V_{tb}
 \sum^8_{i=1}\left[\Ci(\mu){\it P}_{i}+\Cip(\mu){\it
P}_{i}^{\prime}\right].
\eeq
The Wilson coefficients $C_i(\mu)$ and $C_{i}^{\prime}(\mu)$, which are
associated with operators ${\it P}_{i}$ and their chirality--conjugate
partners ${\it P}_{i}^{\prime}$, play the
role of effective coupling constants. In the SM, it is natural to
choose the scale at which the heavy (SM) states decouple at
$\muw\simeq\mw$.  Their values at $\muw$ are
found by imposing matching conditions
between an effective and an underlying theory.
One evaluates the $C_{i}^{(\prime)}$s
at $\mub\sim \mb$ from the
running of the renormalization group equations (RGEs)
from $\muw$ down to $\mub$. 
In order to remove the
dependence on the scale $\mu_b$, at $\muw$ NLO MCs 
need to be imposed~\cite{bmmp94}. 
They have been computed in full
in~\cite{NLOmatchingSM,cdgg98,bmu00}, and we have followed~\cite{gm01}
to apply them here with MCs at $\mub$~\cite{ghw96}.

New physics effects usually appear in the Wilson coefficients
$C_{7,8}$ and $C_{7,8}^{\prime}$
of the 
magnetic and chromo--magnetic operators
$P_{7,8}$ and $P_{7,8}^{\prime}$, respectively.  In the MFV case, as
well as in the SM/2HDM, one usually neglects $C_{7,8}^{\prime}$ as
suppressed by the ratio $\ms/\mb$. This can no longer be done in the
case of GFM, where the $\gluino$ contributions to $C_{7,8}^{\prime}$ are
of a similar strength~\cite{bghw99}.
In SUSY, the new mass scale $\msusy$ is set by $\mgluino$ and the
squark mass $\msquark$, which tend to be heavier than the other
states. (We assume no large mass splittings among the squarks.) It is
reasonable to expect a hierarchy $\mususy\gg\muw$ just like one has
$\muw\gg\mub$. In order for QCD corrections beyond the LO to match
the underlying SUSY theory and the effective theory at $\muw$, we
extend the treatment of~\cite{cdgg98susy,dgg00} to the case of GFM.
The usual procedure, which we follow here, is to assume that
$\mgluino\sim\msquark\sim\mususy$, while all the other SUSY states
play a role at $\muw$~\cite{dgg00}.  We thus do not resum the
logarithms of $m_H/\muw$ and $m_{\wt{q}(\wt{g})}/\mususy$.  The
resummation of QCD correction between $\mususy$ and $\muw$ is
performed with the NLO anomalous dimensions with six quark
flavors~\cite{misiakmunz95}.
We compute two--loop gluon corrections to the $H^{-}$ contributions to
$C_{4,7,8}$ at $\muw$, and those from the $\charginom$, $\neut$ and
$\gluino$ fields to $C_{7,8}^{(\prime)}$ at $\mususy$ as
in~\cite{bmu00}.  
We have verified that their contributions to $C_{1-6}^{(\prime)}$ and
their mixing to $C_{7,8}^{(\prime)}$ are numerically negligible when using
available five--flavor anomalous dimensions~\cite{cmm98}.
We also neglect operators outside of the SM basis.  At LO
they are numerically subdominant~\cite{bghw99},
and we could not identify any enhancement
mechanism beyond the LO.  At the NLO--level, SUSY QCD contributions come
from ${\cal O}(\alpha_s)$ corrections involving the gluino.  For the
states at $\muw$ ($W$, $\phi^-$ and $H^-$), these are absorbed into
the effective vertices by integrating out the $\gluino$
field~\cite{cdgg98susy}. We have calculated these vertices with 
GFM in the down--squark sector~\cite{or2}.

For the fields at $\mususy$, this effective approach cannot be
applied.  Instead of calculating a full set of two--loop diagrams (a
rather formidable task), we include finite threshold corrections to
Yukawa matrices, which become important at large
$\tanb$~\cite{bottomcorr,cgnw00,dgg00}.  As described in detail
in~\cite{or2}, we first integrate out the gluino and the squarks to calculate
effective quark mass matrices in the super-CKM (SCKM) basis. By identifying these with
physical ones, we extract the Yukawa couplings that appear in the higgsino--quark--squark
vertices and in the F--term contributions to squark mass matrices, and
further perturbatively include ``$\tan\beta$
resummation''~\cite{cgnw00,dgg00}, generalized to GFM~\cite{or2} for
self--consistency.

\vspace{0.15cm}
{\it General flavor mixings}.---  In the absence of an underlying
theory of flavor in the squark sector, all the entries of the
$6\times6$ mass matrix square of down--type squarks are in general
nonzero (and likewise for the up--type squarks).
It is natural
to break it into four $3\times3$ submatrices of the $LL$, $LR$, $RL$
and $RR$ sectors. For the 2nd to 3rd generation mixing of
relevance to $\bsgamma$, the departure from the MFV case can be
parameterized by introducing
$\deltadll=
(m^{2}_{d,LL}
)_{23}/{\sqrt{
(m^{2}_{d,LL}
)_{22}
(m^{2}_{d,LL}
)_{33}}}$ and 
$\deltadlr= 
(m^{2}_{d,LR}
)_{23}
  /{\sqrt{
(m^{2}_{d,LL}
)_{33}
(m^{2}_{d,RR}
)_{22}}}$,
and analogously for $\deltadrl$ and $\deltadrr$, where
$m^{2}_{d,LL(RR,LR)}$ are, respectively, the $3\times3$ soft mass matrices
$m^2_{Q(D)}$ and the soft
trilinear term matrix $A_d^{\ast}$ rotated
into the SCKM basis~\cite{or2}.

In the case  
of GFM, these mixings induce new 1--loop
contributions to $\bsgamma$, with one internal line involving
$\gluino$ (or $\neut$) and the other from 
$\sbottom$ turning into 
$\sstrange$
due to the 2nd to 3rd generation mixing. 
We insert the $\delta^d$s at $\mususy$ and all
the corrections described above are generalized to the case of GFM. 
The $6\times6$ mass matrices are diagonalized numerically,
instead of using the less accurate mass--insertion approximation.

\begin{figure}[b!]
\vspace*{-0.2in}
\begin{center}
\begin{minipage}{3in} 
\centerline{
\resizebox*{2.5in}{2.3in}{\includegraphics{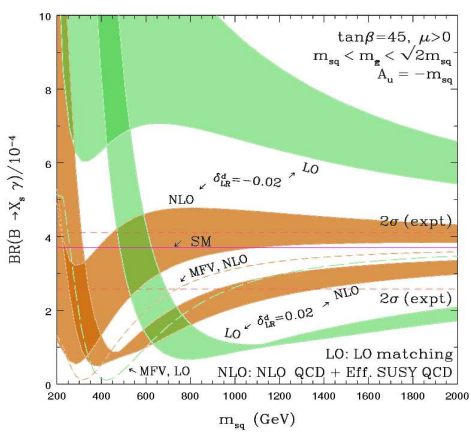}}
}
\end{minipage}
\caption{\label{brmhalft40ldfmup:fig} {\small
(color online).
 We plot $\brbsgamma$
\vs\
$m_{\widetilde{q}}$ 
for $\deltadlr=\pm0.02$ and $0$ (MFV). 
For each case we show the bands obtained by varying
$m^2_{\widetilde{g}}/m^2_{\widetilde{q}}$
in the approximation of using only LO MCs (green/light grey) and of
including NLO corrections (brown/dark grey). 
The SM value and $2\sigma$~C.L.
experimental limits are also marked.
}
}
\end{center}
\end{figure}

\vspace{0.15cm}
{\it Focusing effect}.---  We find that, relative to the LO MC 
at $\muw$, NLO corrections generally reduce SUSY
contributions (mostly from the gluino) to $\brbsgamma$. As a result,
constraints on $\mgluino$
and related SUSY parameters (like the $\chargino$ and $\neut$ soft mass parameters), 
and on the FV couplings become considerably relaxed. This {\em focusing effect}
becomes particularly strong at large $\tanb$. This is illustrated in
Fig.~\ref{brmhalft40ldfmup:fig}. 
We present ranges of the $\brbsgamma$ as a function of 
the common squark mass, $m_{\widetilde{q}}\,\equiv \,\sqrt{(m^2_{Q,U,D})_{ii}}$
in the case of MFV ($\deltadlr=0$) and for
$\deltadlr=\pm 0.02$. We scan over the ranges of
$1<m^2_{\widetilde{g}}/m^2_{\widetilde{q}}<2$
and
apply collider bounds on superpartner masses. The brown (darker) bands
are obtained by using our expressions for dominant NLO--level
contributions, while the green (light) bands correspond to applying the
approximation of the LO matching at $\muw$. One can clearly see a strong
suppression of the SUSY contribution at the NLO--level. While some
focusing is already present in the MFV case, the effect becomes
strongly enhanced in the case of GFM.

\begin{figure}[b!]
\begin{center}
\begin{minipage}{9.2cm} 
\centerline{
\resizebox*{4.6cm}{!}{\includegraphics{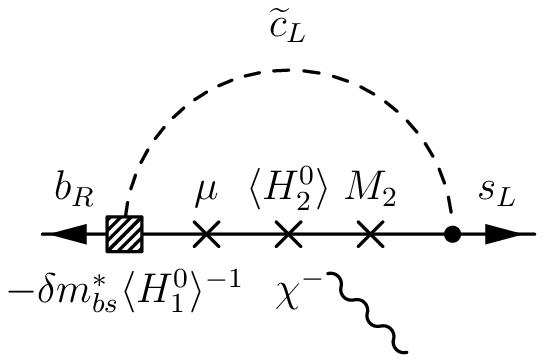}}
\resizebox*{4.6cm}{!}{\includegraphics{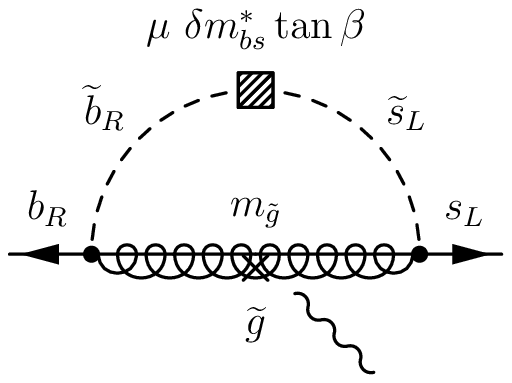}}
}
\end{minipage}
\caption{\label{fdiags:fig} {\small Examples of $\tanb$--enhancement in GFM.
}}
\end{center}
\end{figure}

Focusing comes from two sources. First, RGE evolution between
$\mususy$ and $\muw$ generally reduces SUSY contributions. For example, for
$\msusy\sim1\tev$, 
one finds $\alpha_s(\mususy)/\alpha_s(\muw)\simeq 0.8$, 
$C_7(\muw)\simeq 0.8 C_7(\mususy) +0.06
C_8(\mususy)$ and $C_8(\muw)\simeq 0.8 C_8(\mususy)$.  
(The NLO QCD matching condition at $\mususy$ causes some additional
reduction.)  Second, at $\mu>0$ we find a remarkable tendency of
``alignment'' between the effective $b$--quark mass
and the corresponding penguin operator at large $\tanb$.  This reduces
the gluino contribution to $C_{7,8}$ and, as a 
result, the SUSY contribution to $\brbsgamma$ in the case of GFM.
The essential point is that flavor mixing in the soft
SUSY--breaking terms induces flavor off-diagonal elements in the
effective quark mass matrices through gluino--squark loops.  In
order to rediagonalize them, one also rotates the Yukawa couplings
that appear in the higgsino vertices and squark mass matrices, thus inducing 
${\cal O}(\alpha_s)$ corrections to them in SCKM basis.  The resultant
$C_{7,8}$ is also ``rotated''.
The effect is enhanced by large $\tanb$. 
For example, $\deltadlr$ (or
$\deltadll$) induces a correction  
proportional to 
$\delta m_{bs}^*/\cos \beta$
in the higgsino vertex.
This generates a diagram with an exchange of $\widetilde c_L$ and
$\chi^-$, and another one with the $\widetilde b_R$--$\widetilde s_L$
(with mass insertion 
$\delta m_{bs}^*\mu \tan \beta$
) and the gluino
in the loop, as shown in Fig.~\ref{fdiags:fig}. Both are proportional
to
$\tan \beta$ 
and the rotation occurs in the direction that
reduces the LO effect at $\mu>0$. 
%
If we take $\deltadrr$ (or $\deltadrl$) instead, analogous diagrams
are induced for the $b_L$ to $s_R$ transition, which
reduces the gluino contribution at $\mu>0$ in a similar manner.
The first diagram can be even larger than the LO gluino effect
if the ``diagonal part'' of the chargino penguin 
is arbitrarily large.
For more details see~\cite{or2}.
At $\mu<0$, the rotation described above occurs in the direction
which increases the LO effect.
This competes with the reduction caused by the RGE evolution.
As $\mususy$ increases, the RGE effect increases, while the quark mass
correction decreases.  At $\mususy \sim 1$ TeV, the two effects have
similar size for both signs of $\mu$.

\begin{figure*}[t!]
\onecolumngrid
\vspace*{-0.2in} 
\begin{center}
%
\resizebox{5.5cm}{5.5cm}{\includegraphics{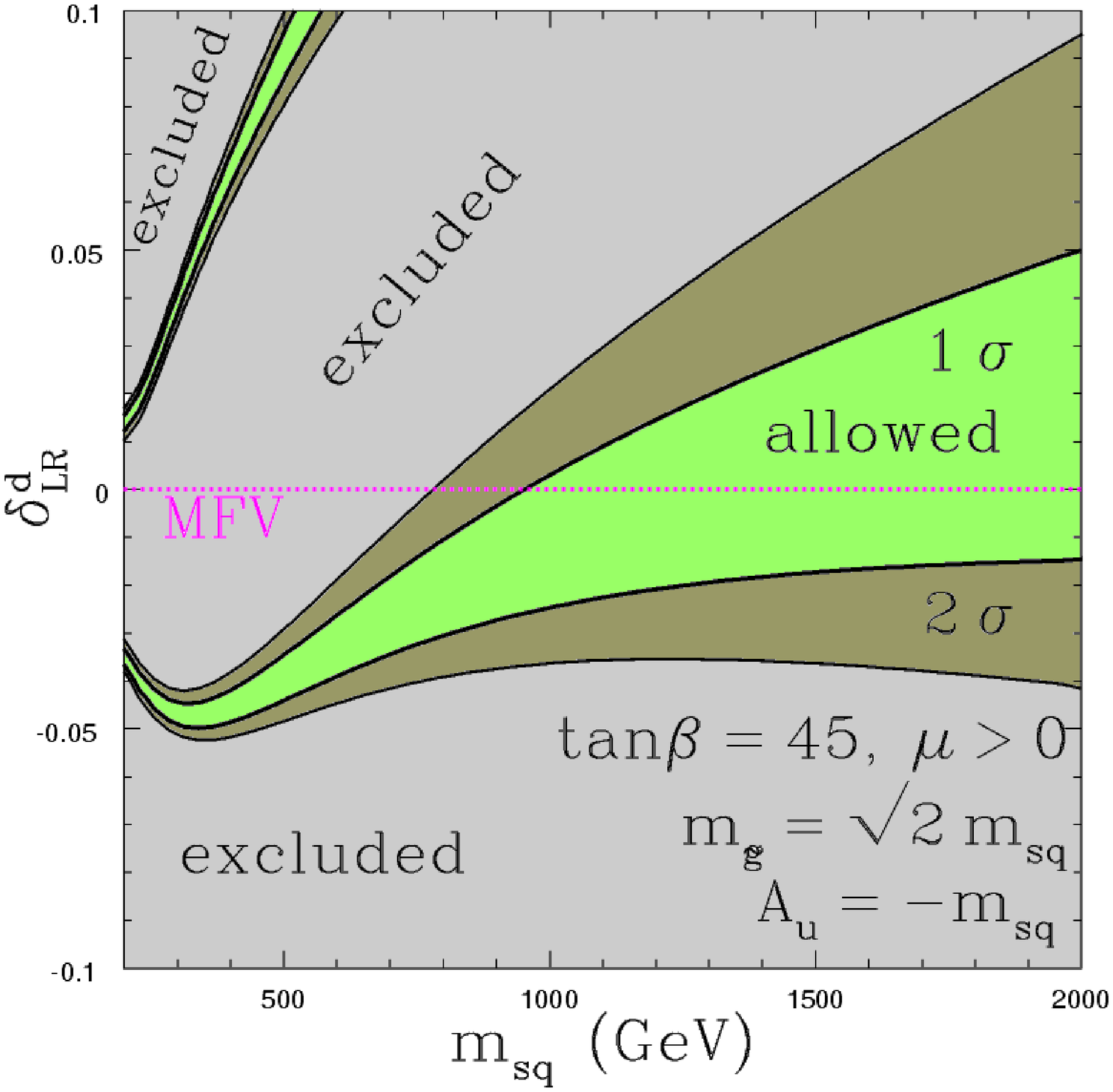}}
\resizebox{5.5cm}{5.5cm}{\includegraphics{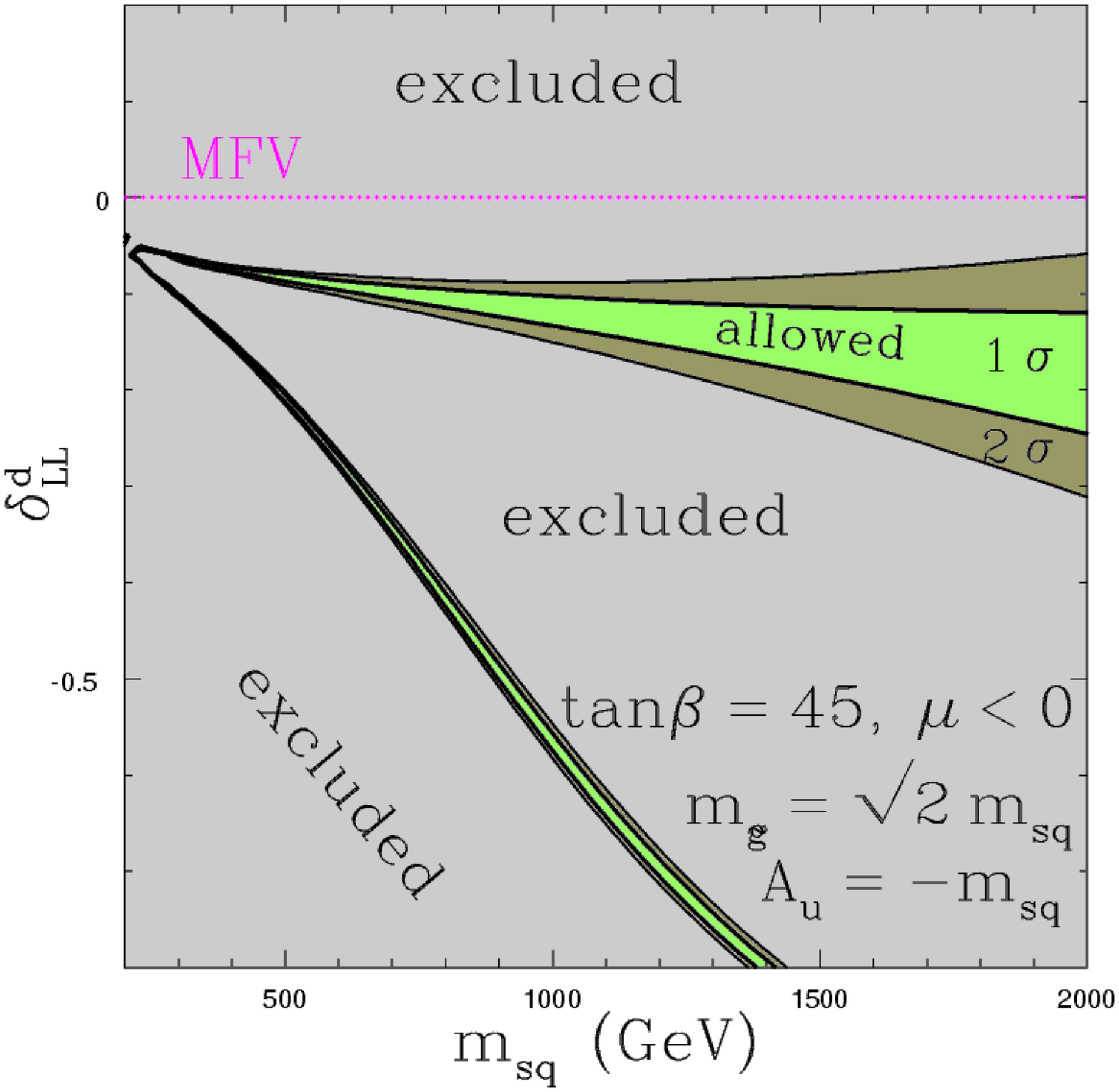}}
\resizebox{5.5cm}{5.5cm}{\includegraphics{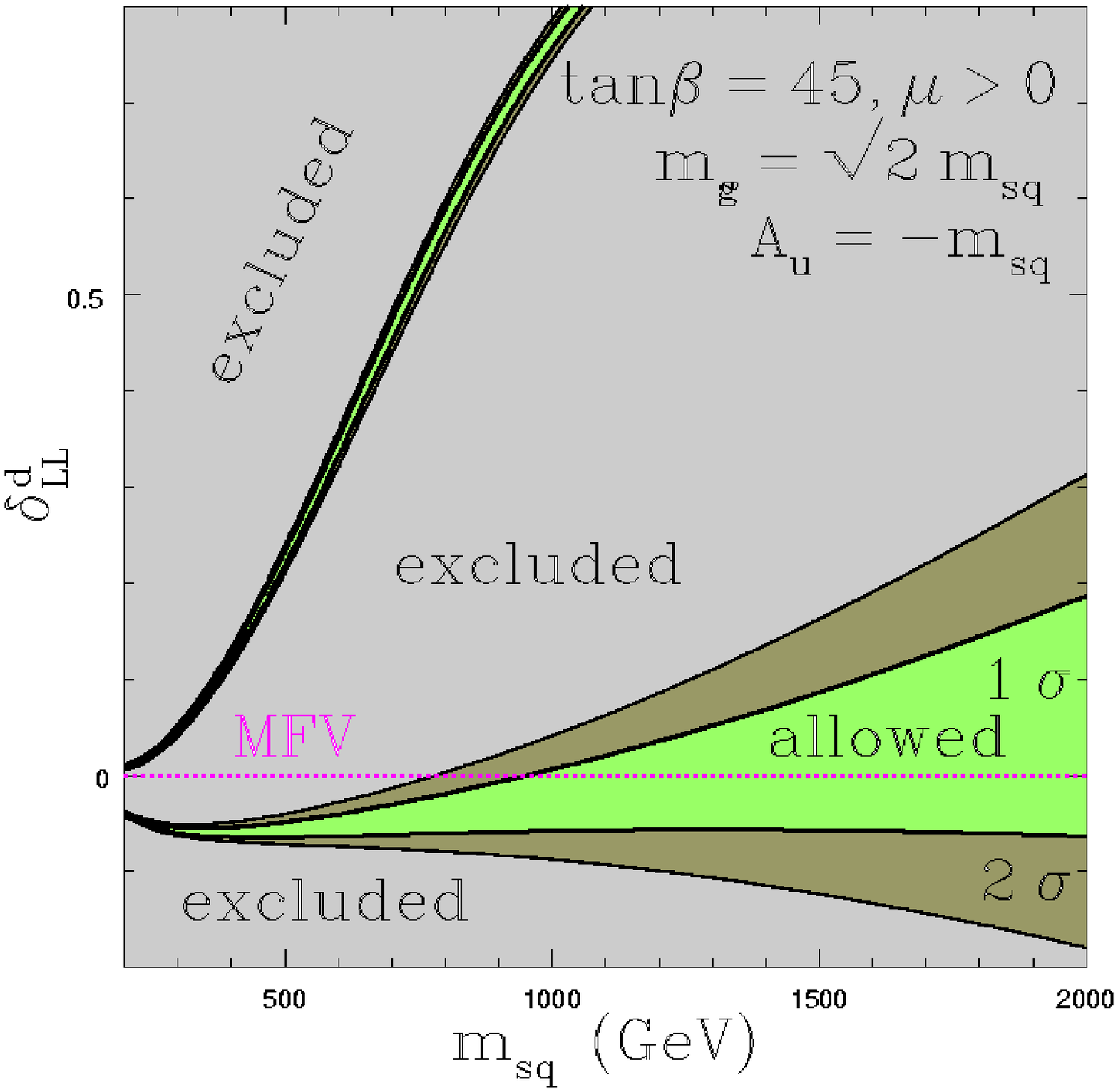}}
\caption{\label{dellrvsmhalf:fig} {\small
(color online). We plot contours of $\brbsgamma$
in the plane of $m_{\widetilde{q}}$
 and $\deltadlr$ (left panel) and $\deltadll$ (
middle and right panels).
In the left and right panels $\mu>0$ and in the middle one $\mu<0$.
Regions (in grey)
beyond the $1\sigma$ (green/light grey) and $2\sigma$~C.L. (dark yellow/grey)
in agreement with
experiment [Eq.~(\ref{bsgexptvalue:ref})] are marked ``excluded''.
}}
\end{center}
\end{figure*}

\vspace{0.15cm}
{\it Relaxation of constraints on SUSY and FV terms}.---
The constraints obtained in the MFV scenario~\cite{cgnw00,dgg00}
are no longer valid with GFM, and their robustness is of particular
phenomenological interest.
We illustrate this by including the above large NLO--level corrections.
In the three panels of
Fig.~\ref{dellrvsmhalf:fig} we show the contours of $\brbsgamma$
 in the plane of
$m_{\widetilde{q}}$
and $\deltadlr$ (left panel) and
$\deltadll$ (middle and right panels) for 
$\tan\beta=45$, $m^2_{\widetilde{g}}/m^2_{\widetilde{q}}=2$,
 $A_u=-m_{\widetilde{q}}$, and $\mu>0$
(left and right panels) and $\mu<0$ (middle panel).
The light green (light grey) and dark yellow (dark grey) bands
agree with the experimental range~(\ref{bsgexptvalue:ref}) at the
$1\sigma$ ($2\sigma$)~C.L. Larger departures are denoted as
``excluded''. In the case of $\deltadlr$, in MFV one finds
$m_{\widetilde{q}}\gsim 800\gev$
 at the $2\sigma$~C.L. for $\mu>0$.
\twocolumngrid
%
%
In contrast, even relatively small nonzero values of
$|\deltadlr|\simeq -0.04$ 
remove the constraint from $\bsgamma$ altogether.  The effect is even
more striking for $\delta^d_{LL}$ and $\mu<0$ where relatively 
light squark becomes again
allowed.  We have checked numerically that, at $\mu>0$, constraints on
FV terms at NLO become also considerably relaxed.  
For example, in Fig.~\ref{dellrvsmhalf:fig} at
$m_{\widetilde{q}}=1000\gev$
 and $\mu>0$,
we find
$\deltadlr(NLO)/\deltadlr(LO)\simeq7$. 
%
Similar GFM deconstraining can also be obtained by introducing
nonzero $\deltadll$s instead, for both $\mu>0$ and $\mu<0$
but with a milder NLO effect.
It is interesting to note the existence of two branches of allowed
solutions in Fig.~\ref{dellrvsmhalf:fig}.  By
simultaneously allowing for more than one $\delta^d$ to become
nonzero, one can relax the constraint from $\bsgamma$ even further,
even at LO~\cite{bghw99}. In addition, in general FV contributions can
also come from the up--squark sector~\cite{up-mixing}. We have checked
that this effect is numerically less important.

Finally, we must reemphasize that in our analysis we assume
$\msusy/\mw$ larger than a few and compute only the leading NLO
effects, which are enhanced by $\msusy/\mw$, large $\tanb$, or flavor
mixing. We also remain in the SM basis of operators, instead of
considering the full operator basis of the MSSM.  Despite this, we do
believe that the results presented here should remain as the dominant
effect in a more complete study.


\vspace{0.15cm}
{\it Conclusions and outlook}.---  
Contributions from new physics to rare processes like $\bsgamma$,
do not necessarily have to be suppressed by the largeness of the
related effective mass scale beyond the SM. Instead, this can be caused by new effects
beyond the LO. In the case of SUSY with GFM, at the
NLO level we have pointed out the existence of one such suppressing effect, which
we call focusing.
Despite this, $\bsgamma$ still maintains strong
sensitivity to even small deviations from MFV
(although less so than at LO)  of
especially the flavor mixing terms $\deltadll$ and $\deltadlr$. This
puts into question the robustness of the commonly assumed constraints
on $\msusy$ in MFV.
It appears that the alignment mechanism is generic to
processes involving chirality flip. We are now
exploring its role in other FCNC processes.

\acknowledgments We thank P.~Gambino, T.~Hurth,
A.~Masiero, M.~Misiak and L.~Silvestrini for helpful comments.
%

\end{document}